\documentclass[12pt,a4paper]{amsart}

\usepackage[T1]{fontenc}
\usepackage[utf8]{inputenc}
\usepackage[english]{babel}
\usepackage{lmodern}
\usepackage{amsmath,amsfonts,amssymb,amscd}
\usepackage{cite,url}
\usepackage[margin=20mm]{geometry}

\numberwithin{equation}{section}
\allowdisplaybreaks
\newcommand{\p}{\partial}
\newcommand{\bsy}{\boldsymbol}
\newcommand{\ds}{\displaystyle}

\title[On the vector Schwarz--KdV equation]
{On the vector Schwarz--KdV equation}

\author{M. Ju. Balakhnev}
\address{M. Ju. Balakhnev, Orel State University,
Komsomolskaja str., 95, 302026, Orel, Russia}
\email{balakhnev@yandex.ru}

\author{V. V. Sokolov}
\address{V. V. Sokolov, Higher School of Modern Mathematics MIPT,
Institutsky lane, 9, 141701, Dolgoprudny, Russia;
National Research University Higher School of Economics,
Usachev str., 6, 119048, Moscow, Russia}
\email{vsokolov@landau.ac.ru}

\thanks{UDC 517.957.}
\thanks{\copyright\ M. Balakhnev, V. Sokolov, 2025.}
\date{Received 31 December 2025}

\begin{document}

\begin{abstract}
A collection of miscellaneous continuous, semi-discrete, and discrete
integrable systems can be associated with each integrable evolution
equation of the KdV type. We give them for the Schwarz--KdV equation and
generalize to the vector case. The existence of these vector
generalizations is a non-trivial experimental fact for which no
mathematical explanation is yet known.
\end{abstract}

\maketitle
\enlargethispage{3pt}

\section{Introduction}

Using the symmetry approach to classification of integrable equations
(see \cite{Sokshab, MikShaSok91, Sok}),
in the papers \cite{SvinSok, MeshSok1} all scalar integrable evolution
equations of the form
\begin{equation}\label{eqev3}
u_t=u_{xxx}+f(u, u_x, u_{xx})
\end{equation}
that have an infinite sequence of local conservation laws were listed.

A generalization of this approach to the case of systems of the form
\begin{equation}\label{eqvec3}
\bsy u_t =\bsy u_{xxx}
+ f_ {2} \, {\bsy u} _ {xx}
+ f_ {1} \, {\bsy u} _ {x}
+ f_ {0} \, {\bsy u}
\end{equation}
where the coefficients $f_i$ depend on scalar products between the
vectors ${\bsy u}, {\bsy u}_{x}, {\bsy u}_{xx}$,
was proposed in \cite{meshsok}.
Some lists of integrable equations \eqref{eqvec3} were obtained in
\cite{meshsok1, balmesh1, balmesh2}.

One of the most beautiful examples of such integrable systems is the
so-called vector Schwarz--KdV equation \cite{SviSok94}:
\begin{equation}\label{S-K}
\bsy u_t =\bsy u_{xxx}
-3\frac{(\bsy u_x,\bsy u_{xx})}{(\bsy u_x,\bsy u_x)}\bsy u_{xx}
+\frac32\frac{(\bsy u_{xx},\bsy u_{xx})}{(\bsy u_x,\bsy u_x)}
\bsy u_x,
\end{equation}
where $\bsy u(x,t)$ is a vector, which belongs to an $N$-dimensional
vector space $V$ equipped with a scalar product $(\cdot,\cdot)$,
$\bsy u_t=\p \bsy u/\p t,
\bsy u_i=\p^i \bsy u/\p x^i$.
In component form it is a system of $N$ evolution equations invariant
with respect to the orthogonal group.

Any integrable equation \eqref{eqev3} generates several associated
differential, difference-differential and fully discrete integrable
equations \cite{Adler2,Adler3}. In Section \ref{scalar} we describe
these integrable models in the case of the scalar Schwarz--KdV equation
\begin{equation}\label{SKdV}
u_t=u_{xxx}-\frac32 \frac{u_{xx}^2}{u_x}.
\end{equation}

The main question, which we are discussing in this paper, is whether
there exists such a variety of associated integrable vector equations
for each vector equation \eqref{eqvec3}. The answer is not a priori
obvious. We will show that this is the case for the vector equation
\eqref{S-K}. While in the scalar case the search for related integrable
systems is reduced to rather simple algorithmizable calculations,
in the vector case the calculations become complex and sometimes
require non-trivial tricks.

\section{Scalar case}\label{scalar}

Let us demonstrate the collection of associated integrable systems
for the scalar Schwarz--KdV equation \eqref{SKdV}.

\subsection{B\"acklund transformation.}

We assume that each function
$u_n(x,t), \, n\in Z$ satisfies equation \eqref{SKdV}.
Then the formula
\begin{equation}\label{back}
(u_{n+1})_x=
\alpha \frac{(u_{n+1}-u_{n})^2}{(u_{n})_x},
\qquad \alpha\in \mathbb C
\end{equation}
defines an auto-B\"acklund transformation \cite{Adler_98}
for \eqref{SKdV}. This semi-discrete chain is itself an integrable
model. In the paper \cite{vesshab} such chains related to equations
\eqref{eqev3} are called {\it dressing chains}.

To obtain formula \eqref{back}, we write
\begin{equation}\label{backanz}
(u_{n+1})_x = Q\Big((u_{n})_x, u_{n}, u_{n+1}\Big),
\end{equation}
differentiate it with respect to $t$, and eliminate all $t$-derivatives
using \eqref{SKdV} and all $x$-derivatives of $u_{n+1}$ by virtue of
\eqref{backanz}.
The remaining variables
$u_{n+1}, u_n, (u_{n})_x, \dots , (u_{n})_{xxxx}$
are regarded as independent jet variables. Splitting with respect to
$(u_{n})_{xx}, (u_{n})_{xxx},(u_{n})_{xxxx}$,
we obtain an overdetermined system of non-linear PDEs for
$Q\Big((u_{n})_x, u_{n}, u_{n+1}\Big)$.
This can be solved quite easily. The chain \eqref{back} corresponds
to the ``most interesting'' solution of this system.
The existence of other solutions can be explained by the fact that
equation \eqref{SKdV} is invariant with respect to the
M\"obius transformations
$$
\tilde u = \frac{a u+b}{c u+d},
\qquad a,b,c,d\in \mathbb C.
$$

\subsection{Volterra-type chain.}

The pair of equations \eqref{SKdV} and \eqref{back} is compatible
with the following chain of Volterra type\footnote{Such integrable
chains were classified in \cite{Yamilov}.}:
\begin{equation}\label{volt}
(u_n)_z=
\beta \frac{(u_{n+1}-u_{n})(u_{n-1}-u_{n})}{u_{n+1}-u_{n-1}},
\qquad \beta\in \mathbb C.
\end{equation}
Compatibility provides the existence of solutions $u(n,x,t,z)$
satisfying \eqref{SKdV}.

The chain \eqref{back} can be regarded as a hyperbolic equation,
where the variable $x$ is continuous and $n$ is discrete.
A usual integrable hyperbolic equation $u_{xy}=F$ has integrable
evolution symmetries with respect to both variables $x$ and $y$.
The equation \eqref{back} has the $x$-symmetry \eqref{SKdV}
and $n$-symmetry \eqref{volt}.

\subsection{Negative flow.}

Now, let us consider the relation \eqref{back}, its shift
$$
(u_{n})_x=
\alpha \frac{(u_{n}-u_{n-1})^2}{(u_{n-1})_x},
$$
and \eqref{volt} together with its $x$-derivative.
Expressing
$(u_{n+1})_x,(u_{n-1})_x,u_{n+1},u_{n-1}$
in terms of
$u_n, (u_{n})_x, (u_{n})_z, (u_{n})_{zx}$
from these four equations and differentiating \eqref{volt}
in $x$ twice, we obtain the equation
\begin{equation}\label{3ord}
u_{zxx} =
\frac{u_{zx}^2-\beta^2 u_x^2}{2 u_z}
+ \frac{u_{xx} u_{zx}}{u_x} + 2 \alpha u_z,
\end{equation}
where $u = u_n. $
This equation is a negative flow for \eqref{SKdV}.
Moreover, a recursion operator for \eqref{SKdV} can be extracted
from \eqref{3ord}, see \cite{Adler2,Adler3}.

\subsection{System of NLS type.}

Equation \eqref{SKdV} is an $x$-symmetry for \eqref{3ord}.
But equation \eqref{3ord} turns out to be really integrable
and therefore has $z$-symmetries
(cf. \cite{AdlerShabat2012}) of the form
$$
u_{\tau}=F_1(u,u_x, u_z, u_{zx},u_{zz},..),
\qquad
(u_x)_{\tau}=F_2(u,u_x, u_z, u_{zx}, u_{zz},..).
$$
Computation shows that the simplest $z$-symmetry has the form
$$
F_1=
-\frac{u_{zzx}u_z}{u_{x}}
+\frac{u_{zz}u_{zx}}{u_{x}}
+\frac{1}{2}\frac{u_{z}u_{zx}^2}{u_{x}^2},
\qquad F_2=(F_1)_x.
$$

Denoting $u_x$ by $v$, we obtain the following second-order system:
\begin{equation}\label{separ}
u_\tau=
-\frac{v_{zz}u_z}{v}
+\frac{v_z u_{zz}}{v}
+\frac{1}{2}\frac{u_z v_z^2}{v^2},
\qquad
v_\tau=
-\frac{v_{zz}v_z}{v}
+\frac{u_{zz}}{vu_z}\left(v_z^2-\beta^2v^2\right)
+\frac{1}{2}\frac{v_z^3}{v^2}
+\beta^2 v_z.
\end{equation}

The differential substitution
$$
\tau = \beta^{-1} y+\frac{1}{2}z,\qquad
u_z=
e^{\sqrt{\beta} u'-\sqrt{\beta} v'/2}
\Big(\beta e^{\sqrt{\beta} v'}
+(e^{\sqrt{\beta} v'})_z\Big),
\qquad v=e^{\sqrt{\beta} v'}
$$
reduces \eqref{separ} to the potential derivative NS-system
\begin{equation*}
u_y= u_{zz}+v_z u_z^2-\frac14 \beta v_z,\qquad
v_y=-v_{zz}+u_z v_z^2-\beta u_z.
\end{equation*}
Note that it is integrable \cite{MikhShab87}
(a special case of system w3) for any constant coefficients
of $v_z$ and $u_z$.

If desired, we can continue to create a net of integrable
systems generated by the Schwarz--KdV equation.
In particular, for the system
\begin{equation}\label{NLS1}
u_y= u_{zz}+v_z u_z^2+k_1 v_z,\qquad
v_y=-v_{zz}+u_z v_z^2+k_2 u_z.
\end{equation}
one can find the following B\"acklund transformation:
\begin{align*}
U_z&=f u_z-(1+f^{-1})f_v,\qquad \quad
V_z=f^{-1}\big(v_z-(1+f^{-1})f_u\big),
\end{align*}
where the function $f(u,v,U,V)$ is defined by the compatible
system of ODEs
$$
f_U=f_u,\quad f_V=f_v,\quad
f_{vv}=
-\frac{1}{2}
\frac{k_1(f-1)f^2-(f+3)f_v^2}{f(f+1)},\quad
f_{uu}=
-\frac{1}{2}
\frac{k_2(f-1)f^2-(f+3)f_u^2}{f(f+1)},
$$
which is a semi-discrete system for two fields $u$ and $v$.
However, in this paper we do not aim to describe in some sense
a complete web of integrable systems that can be obtained
from the Schwarz--KdV equation.

\subsection{Superposition formula}

We assume now that functions
$
u(x,t, z_1,\dots, z_k, n_1,\dots n_k)
$
satisfy equation \eqref{SKdV}, equation \eqref{back}
with different parameters $\alpha_i$ and equation \eqref{volt}
with different parameters $\beta_i$.
Let us denote by $T_i$ the shift $n_i \mapsto n_{i+1}$. Then
$$
T_i(u_x)=
\alpha_i \frac{\big(T_i(u)-u\big)^2}{u_x}.
$$
Consider the relations
$$
u_xT_i(u_x)=\alpha_i\big(T_i(u)-u\big)^2,\qquad
T_j(u_x)T_iT_j(u_x)=
\alpha_i\big(T_iT_j(u)-T_j(u)\big)^2,
$$
$$
u_xT_j(u_x)=\alpha_j\big(T_j(u)-u\big)^2,\qquad
T_i(u_x)T_jT_i(u_x)=
\alpha_j\big(T_jT_i(u)-T_i(u)\big)^2.
$$
Eliminating $x$-derivatives from these equations and
using that $T_iT_j(u)=T_jT_i(u)$, we obtain the relation
$$
\alpha_j^2
\Big(T_iT_j(u)-T_i(u)\Big)^2
\Big(T_j(u)-u\Big)^2
=
\alpha_i^2
\Big(T_iT_j(u)-T_j(u)\Big)^2
\Big(T_i(u)-u\Big)^2,
$$
Factorizing this, we obtain two candidates for the superposition
formula. It can be verified that only the relation
\begin{equation}\label{supfor}
\alpha_j
\Big(T_iT_j(u)-T_i(u)\Big)
\Big(T_j(u)-u\Big)
=
\alpha_i
\Big(T_iT_j(u)-T_j(u)\Big)
\Big(T_i(u)-u\Big),
\end{equation}
is invariant with respect to the $x$-derivation.
This superposition formula for the equation \eqref{SKdV}
can be represented by the commutative diagram
\begin{equation*}
\begin{CD}
T_i(u) @>\alpha_j>> T_iT_j(u)\\
@A\alpha_i AA @AA\alpha_i A\\
u @>>\alpha_j> T_j(u)
\end{CD}\,.
\end{equation*}
It is well known \cite{ABobSur} that this integrable discrete
equation satisfies the $3d$-consistency condition.

\subsection{3-D consistent hyperbolic system \cite{Adler2}.}

Let us extend the system
\begin{equation}\label{3ords}
u_{z_{i}xx}=
\frac{u_{z_{i}x}^2-\beta_i^2u_x^2}{2u_{z_i}}
+\frac{u_{xx}u_{z_i x}}{u_x}
+2\alpha_i u_{z_i},
\qquad i=1,\dots,k
\end{equation}
of equations \eqref{3ord} by
\begin{equation}\label{2ords}
u_{z_i z_j}=
\frac{
\alpha_i u_{z_i}u_{x z_j}
-\alpha_j u_{z_j}u_{x z_i}
}{
(\alpha_i-\alpha_j)u_x
},
\qquad i\ne j,\quad i,j=1,\dots,k.
\end{equation}
One can verify that the system \eqref{3ords}, \eqref{2ords}
is compatible:
$$
\Big(u_{z_i xx}\Big)_{z_j}
=
\Big(u_{z_j xx}\Big)_{z_i}
=
\Big(u_{z_j z_i}\Big)_{xx},
\qquad
\Big(u_{z_i z_j}\Big)_{z_k}
=
\Big(u_{z_j z_k}\Big)_{z_i}
=
\Big(u_{z_k z_i}\Big)_{z_j},
$$
where the derivatives are calculated by virtue of
\eqref{3ords}, \eqref{2ords}.
This means that there exists a common solution
$u(x,z_1,\dots,z_k)$ of this system.
For a given equation \eqref{3ords} equations \eqref{2ords}
were found from the compatibility conditions by some non-trivial
computation. It is interesting to note that \eqref{2ords}
is compatible on its own, i.e. without using consequences
of system \eqref{3ords}.

Some of the equations from Sections 2.1-2.6 are well known.
The others were found by more or less simple calculations
by the algorithms described in Adler's papers.
In contrast with this, finding similar objects for a vector
Schwarz--KdV equation is not yet algorithmizable, requires
various non-trivial tricks, and is sometimes extremely time
and memory-consuming.

\section{Around the vector Schwarz--KdV equation}

In this section, for aesthetic reasons, instead of a single
variable $\bsy u$ with subscripts, we use various bold characters.
The transition to index notation in the vector formulas is easily
performed by comparing with the corresponding formulas from
Section 2, taking into account that scalar formulas should coincide
with vector ones if $N=1.$

\subsection{B\"acklund transformation.}

For vector equations \eqref{eqvec3} the B\"acklund transformations
we are dealing with have the form
$$
\bsy u_x=K_1\bsy v_x+K_2\bsy u+K_3\bsy v,
$$
where $K_i$ are some functions of the six scalar products
of vector variables $\bsy v_x,\bsy u,\bsy v$.
In the paper \cite{meshsok1}, the following B\"acklund transformation
for the vector Schwarz--KdV equation \eqref{S-K} was found:
\begin{equation}\label{vecSKdVBT}
\bsy u_x=
\frac{\mu}{\bsy v_x^2}
(\bsy u-\bsy v,\bsy v_x)(\bsy u-\bsy v)
-\frac{\mu}{2\bsy v_x^2}
(\bsy u-\bsy v)^2\bsy v_x.
\end{equation}
It is easy to verify that \eqref{vecSKdVBT} coincides with
\eqref{back} if $N=1$,
$\bsy u\to u_{n+1},\bsy v\to u_n,\mu=2\alpha$.

At first glance, the vectors $\bsy u$ and $\bsy v$ are not
included in the formula \eqref{vecSKdVBT} symmetrically.
However, multiplying \eqref{vecSKdVBT} by
$\bsy v,\bsy u,\bsy v_x$ and $\bsy u_x$ and expressing scalar
products $(\bsy u,\bsy v_x)$, $(\bsy v,\bsy v_x)$,
$(\bsy u_x,\bsy v_x)$ and $\bsy v_x^2$ in terms of the
remaining ones, it is not difficult to check that
$$
\bsy v_x=
\frac{\mu}{\bsy u_x^2}
(\bsy v-\bsy u,\bsy u_x)(\bsy v-\bsy u)
-\frac{\mu}{2\bsy u_x^2}
(\bsy v-\bsy u)^2\bsy u_x.
$$

\subsection{Volterra-type chain.}

The vector Volterra-type chain has the form
\begin{equation}\label{anzVol}
\bsy v_z=Q_1\bsy u+Q_2\bsy v+Q_3\bsy w,
\end{equation}
where the coefficients $Q_i$ are functions of all scalar
products between the vectors $\bsy u,\bsy v,\bsy w$.
This chain has to be compatible with equation \eqref{S-K}
for $\bsy u,\bsy v$, and $\bsy w$ modulo equation
\eqref{vecSKdVBT} taken together with
\begin{equation}\label{back2}
\bsy w_x=
\frac{\mu}{\bsy v_x^2}
(\bsy w-\bsy v,\bsy v_x)(\bsy w-\bsy v)
-\frac{\mu}{2\bsy v_x^2}
(\bsy w-\bsy v)^2\bsy v_x.
\end{equation}

The compatibility condition
\begin{equation}\label{comty}
(Q_1\bsy u+Q_2\bsy v+Q_3\bsy w)_t
=
\left(
\bsy v_{xxx}
-3\frac{(\bsy v_x,\bsy v_{xx})}{(\bsy v_x,\bsy v_x)}
\bsy v_{xx}
+\frac32\frac{(\bsy v_{xx},\bsy v_{xx})}{(\bsy v_x,\bsy v_x)}
\bsy v_x
\right)_z
\end{equation}
involves vectors $\bsy u,\bsy v,\bsy w,$ whose $x$-derivatives
are related by \eqref{vecSKdVBT} and \eqref{back2}.
It is easy to see that the vector $\bsy v_{xxx}$ in the
relation \eqref{comty} cancels out, and we can take
\begin{equation}\label{vecvar}
\bsy u,\quad \bsy v,\quad \bsy w,\quad
\bsy v_x,\quad \bsy v_{xx}
\end{equation}
as independent vector jet variables.
Finding the scalar products of relations \eqref{vecSKdVBT}
and \eqref{back2} with independent vectors, we obtain a set
of algebraic relations between scalar products of vectors
\eqref{vecvar}. Each of these scalar products can be expressed
in terms of products between vectors \eqref{vecvar}.
The latter products we regard as independent.

Splitting \eqref{comty} with respect to the independent vectors,
we obtain a system of five PDEs for functions $Q_i$.
Each of these PDEs can be split with respect to the independent
scalar products that depend on $x$-derivatives.
Solving the resulting system of linear PDEs for the functions
$Q_i$, each of which depends on six variables, we have found
the following vector Volterra-type chain
\begin{equation}\label{V3Rn1}
\bsy v_z=
\frac{
(\bsy w-\bsy v)^2(\bsy u-\bsy v)
-(\bsy u-\bsy v)^2(\bsy w-\bsy v)
}{
(\bsy u-\bsy w)^2
}.
\end{equation}
It can be verified that \eqref{V3Rn1} is compatible with
\eqref{vecSKdVBT}.
If $N=1$, then \eqref{V3Rn1} coincides with \eqref{volt}
after the replacement
$\{\bsy v,\bsy u,\bsy w\}\to\{u_n,u_{n-1},u_{n+1}\}$.
Under the additional assumption
$|\bsy v|=|\bsy u|=|\bsy w|=1$
this integrable vector Volterra-type chain was found in
\cite{Adler} (see formula $V_3$).

\subsection{Negative flow.}

Let us assume that equations \eqref{V3Rn1}, \eqref{vecSKdVBT},
and \eqref{back2} hold.
We are looking for a relation of the form
\begin{equation}\label{neg_f}
\bsy v_{zxx}
=a_1\bsy v_{zx}+a_2\bsy v_{xx}
+a_3\bsy v_z+a_4\bsy v_x+a_5\bsy v,
\end{equation}
where the coefficients depend on the scalar products of vectors
\begin{equation}\label{newjetvec}
\bsy v_{zx},\quad \bsy v_{xx},\quad
\bsy v_z,\quad \bsy v_x,\quad \bsy v.
\end{equation}
It is clear that $\bsy v_z,\bsy v_{zx},\bsy v_{zxx}$
are linear combinations of
\begin{equation}\label{oldjetvec}
\bsy u,\quad \bsy v,\quad \bsy w,\quad
\bsy v_x,\quad \bsy v_{xx}.
\end{equation}
Using these relations, we can express $\bsy u,\bsy w$
through $\bsy v_{zx},\bsy v_{xx},\bsy v_z,\bsy v_x,\bsy v$,
and obtain a relation of the form \eqref{neg_f}.
The main problem is that the coefficients in this relation
depend on the scalar products of vectors \eqref{oldjetvec}
and we have to express these scalar products through the
scalar products of vectors \eqref{newjetvec}.
These two collections of scalar products are related by
a huge system of non-linear algebraic equations, which can
be explicitly solved. As a result of this computation we obtain
\begin{equation}\label{coeff}
\begin{array}{c}
\ds
a_1=\Big(\ln(\bsy v_z,\bsy v_x)\Big)_x,
\qquad
a_2=\frac12\Big(\ln\bsy v_x^2\Big)_z,
\qquad a_5=0,
\\[6mm]
\ds
a_3=\mu
-\frac{(\bsy v_{zx},\bsy v_{xx})}{(\bsy v_z,\bsy v_x)}
-\frac{(\bsy v_{zx},\bsy v_x)^2}{2(\bsy v_z,\bsy v_x)^2}
-\frac{(\bsy v_{zx},\bsy v_z)^2\bsy v_x^2}
{2(\bsy v_z,\bsy v_x)^2\bsy v_z^2}
+\frac{\bsy v_{zx}^2\bsy v_x^2}{2(\bsy v_z,\bsy v_x)^2}
+\frac{
(\bsy v_{zx},\bsy v_x)(\bsy v_{xx},\bsy v_x)
}{
(\bsy v_z,\bsy v_x)\bsy v_x^2
}
+\frac{\bsy v_x^2}{2\bsy v_z^2},
\\[6mm]
\ds
a_4=
\frac{(\bsy v_{zx},\bsy v_z)^2}
{(\bsy v_z,\bsy v_x)\bsy v_z^2}
-\frac{
(\bsy v_{xx},\bsy v_z)(\bsy v_{zx},\bsy v_x)
}{
(\bsy v_z,\bsy v_x)\bsy v_x^2
}
-\frac{\bsy v_{zx}^2}{(\bsy v_z,\bsy v_x)}
-\frac{(\bsy v_z,\bsy v_x)}{\bsy v_z^2}.
\end{array}
\end{equation}

In the scalar limit $N=1$ formula (\ref{neg_f}) gives us
(\ref{3ord}) after the replacement
$\bsy v\to u,\mu=2\alpha$ and $\beta=1$.

\subsection{System of NLS type.}

The simplest $z$-symmetry of (\ref{neg_f}), \eqref{coeff}
has the following form:
\begin{equation}\label{z-sym}
\begin{aligned}
\bsy u_\tau&=
-\frac{\bsy u_z^2}{(\bsy u_z,\bsy u_x)}\bsy u_{zzx}
+\frac{(\bsy u_z,\bsy u_{zx})}{(\bsy u_z,\bsy u_x)}
\bsy u_{zz}
-\bsy u_z^2
\Big(\frac{1}{(\bsy u_z,\bsy u_x)}\Big)_z
\bsy u_{zx}+\\
&+\left(
\frac{\bsy u_z^2(\bsy u_x,\bsy u_{zx})_z}
{\bsy u_x^2(\bsy u_z,\bsy u_x)}
-\frac{
\bsy u_z^2(\bsy u_x,\bsy u_{zx})(\bsy u_x,\bsy u_z)_z
}{
\bsy u_x^2(\bsy u_z,\bsy u_x)^2
}
-\frac{\bsy u_z^2(\bsy u_x,\bsy u_{zx})^2}
{\bsy u_x^4(\bsy u_z,\bsy u_x)}
\right)\bsy u_x+\\
&+\left(
\frac{
(\bsy u_z,\bsy u_{zzx})-(\bsy u_{zx},\bsy u_{zz})
}{
(\bsy u_z,\bsy u_x)
}
-2\frac{(\bsy u_x,\bsy u_{zx})_z}{\bsy u_x^2}
+2\frac{(\bsy u_x,\bsy u_{zx})^2}{\bsy u_x^4}
-\frac32\frac{(\bsy u_z,\bsy u_{zx})^2}
{(\bsy u_z,\bsy u_x)^2}
-\right.\\
&\left.
-\frac{
(\bsy u_z,\bsy u_{zx})(\bsy u_x,\bsy u_{zz})
}{
(\bsy u_z,\bsy u_x)^2
}
+\frac12\frac{\bsy u_z^2\bsy u_{zx}^2}
{(\bsy u_z,\bsy u_x)^2}
+2\frac{
(\bsy u_x,\bsy u_{zx})(\bsy u_z,\bsy u_x)_z
}{
\bsy u_x^2(\bsy u_z,\bsy u_x)
}
-\frac12\frac{
\bsy u_z^2(\bsy u_x,\bsy u_{zx})^2
}{
\bsy u_x^2(\bsy u_z,\bsy u_x)^2
}
\right)\bsy u_z.
\end{aligned}
\end{equation}
If $N=1$, this symmetry coincides with the symmetry from
Section 2.4. As in the scalar case, we can rewrite
(\ref{z-sym}) as a system of equations of the form
$$
\bsy u_\tau=
F_1(\bsy u_z,\bsy u_{zz},\bsy v,\bsy v_z,\bsy v_{zz}),
\qquad
\bsy v_\tau=
F_2(\bsy u_z,\bsy u_{zz},\bsy v,\bsy v_z,\bsy v_{zz}),
$$
where $\bsy v=\bsy u_x$.
After the transformation
$$
\bsy u=
\frac{\bsy v'_z}{\xi}
+\frac{(\bsy v',\bsy v'_z)\bsy u'}{\xi f g}
+\frac{1}{2\xi}
\Big(f-(f\bsy u'+g\bsy v',\bsy v'_z)\xi^{-2}\Big)
\Big(\frac{\bsy u'}{g}+\frac{\bsy v'}{f}\Big),
\qquad \bsy v=\bsy v',
$$
where
$\xi^2=(\bsy u',\bsy v')+fg,\ f=|\bsy v'|,\ g=|\bsy u'|$
we obtain a constant-separant system of the form
\begin{equation}\label{sysUV}
\bsy u'_\tau=
\bsy u'_{zz}
+\Phi_1(\bsy u',\bsy u'_z,\bsy v',\bsy v'_z),
\qquad
\bsy v'_\tau=
-\bsy v'_{zz}
+\Phi_2(\bsy u',\bsy u'_z,\bsy v',\bsy v'_z).
\end{equation}
In the scalar limit this system is reduced to
$$
u_\tau=
u_{zz}+\frac{(u^2-u_z^2)v_z}{2uv}
-\frac{u_z^2}{u}+\frac{u_z}{2},
\qquad
v_\tau=
-v_{zz}+\frac{(v^2-v_z^2)u_z}{2uv}
+\frac{v_z^2}{v}+\frac{v_z}{2}.
$$
The terms $\frac12u_z$ and $\frac12v_z$ in the right hand sides
can be eliminated by the Galilean transformation and then
using the transformation $u=e^{a u'},\,v=e^{-2v'/a},$
we obtain the system (\ref{NLS1}) with
$k_1=-a^{-2},\,k_2=-a^2/4$ for $(u',v')$.

The explicit form of the system \eqref{sysUV} is very
unpresentable. After restriction $\bsy u'^2=\bsy v'^2=1$
to the sphere it becomes more compact:
\begin{equation}\label{sphUV}
\begin{aligned}
\bsy u_\tau&=\bsy u_{zz}
+4\left(
\frac{(\bsy u,\bsy v_z)(\bsy u_z,\bsy v)}{\bsy w^4}
-\frac{(\bsy u_z,\bsy v_z)}{\bsy w^2}
-\frac14(\ln\bsy w^2)_z
\right)\bsy u_z\\
&+2\left(
\frac{\bsy u_z^2}{\bsy w^2}
-\frac{(\bsy w+\bsy u_z,\bsy v)^2}{\bsy w^4}
\right)\bsy v_z
+4\left(
\frac{(\bsy u,\bsy v_z)(\bsy w+\bsy u_z,\bsy v)^2}
{\bsy w^6}
+\frac{\bsy u_z^2(\bsy w-\bsy v_z,\bsy u)}{\bsy w^4}
\right)\bsy w,\\
\bsy v_\tau&=-\bsy v_{zz}
+4\left(
\frac{(\bsy u,\bsy v_z)(\bsy u_z,\bsy v)}{\bsy w^4}
-\frac{(\bsy u_z,\bsy v_z)}{\bsy w^2}
+\frac14(\ln\bsy w^2)_z
\right)\bsy v_z+\\
&+2\left(
\frac{\bsy v_z^2}{\bsy w^2}
-\frac{(\bsy w-\bsy v_z,\bsy u)^2}{\bsy w^4}
\right)\bsy u_z
+4\left(
\frac{(\bsy u_z,\bsy v)(\bsy w-\bsy v_z,\bsy u)^2}
{\bsy w^6}
-\frac{\bsy v_z^2(\bsy w+\bsy u_z,\bsy v)}{\bsy w^4}
\right)\bsy w,
\end{aligned}
\end{equation}
where $\bsy w=\bsy u+\bsy v$ and the terms
$\frac12\bsy u_z$ and $\frac12\bsy v_z$
are omitted as trivial.

\medskip

\subsection{3-D consistent hyperbolic system.}

The vector analog of equations \eqref{3ords},
which satisfy the $3D$ consistency condition, has the form
$$
\bsy u_{z_1z_2}
=
b_1\bsy u_{x z_1}
+b_2\bsy u_{x z_2}
+b_3\bsy u_{z_1}
+b_4\bsy u_{z_2}
+b_5\bsy u_x,
$$
where
$$
b_1=
\frac{\alpha_2}{(\alpha_2-\alpha_1)}
\frac{(\bsy u_{z_1},\bsy u_{z_2})}
{(\bsy u_{z_1},\bsy u_x)},
\qquad
b_2=
\frac{\alpha_1}{(\alpha_1-\alpha_2)}
\frac{(\bsy u_{z_1},\bsy u_{z_2})}
{(\bsy u_{z_2},\bsy u_x)},
$$
$$
b_3=
\frac{
(\bsy u_{x z_2},\bsy u_x)
(\bsy u_{z_1},\bsy u_x)\alpha_1
+\big(
(\bsy u_x,\bsy u_x)(\bsy u_{z_2},\bsy u_{x z_1})
-(\bsy u_{xz_1},\bsy u_x)(\bsy u_{z_2},\bsy u_x)
\big)\alpha_2
}{
(\bsy u_x,\bsy u_x)
(\bsy u_{z_1},\bsy u_x)(\alpha_1-\alpha_2)
},
$$
$$
b_4=
\frac{
(\bsy u_{x z_1},\bsy u_x)
(\bsy u_{z_2},\bsy u_x)\alpha_2
+\big(
(\bsy u_x,\bsy u_x)(\bsy u_{z_1},\bsy u_{x z_2})
-(\bsy u_{x z_2},\bsy u_x)(\bsy u_{z_1},\bsy u_x)
\big)\alpha_1
}{
(\bsy u_x,\bsy u_x)
(\bsy u_{z_2},\bsy u_x)(\alpha_2-\alpha_1)
},
$$
$$
b_5=
\frac{
(\bsy u_{z_1},\bsy u_{z_2})
\big(
(\bsy u_{x z_2},\bsy u_x)
(\bsy u_{z_1},\bsy u_x)\alpha_1
-(\bsy u_{x z_1},\bsy u_x)
(\bsy u_{z_2},\bsy u_x)\alpha_2
\big)
}{
(\bsy u_x,\bsy u_x)
(\bsy u_{z_1},\bsy u_x)
(\bsy u_{z_2},\bsy u_x)(\alpha_2-\alpha_1)
}.
$$
The calculations for finding these coefficients are extremely
difficult, and the intermediate formulas contain several hundred
thousand terms. The resulting equations are compatible by virtue
of the equations \eqref{neg_f}, \eqref{coeff}.
Unlike the scalar case, they are incompatible without the
differential consequences of \eqref{neg_f}, \eqref{coeff}.

\subsection{Vector superposition formula.}

Assume that the following identities hold:
\begin{equation}\label{4bvec}
\begin{aligned}
\bsy u_x&=\Phi(\bsy u,\bsy v,\bsy v_x,\mu),\quad
\bsy w_x=\Phi(\bsy w,\bsy v,\bsy v_x,\nu),\\
\bsy s_x&=\Phi(\bsy s,\bsy u,\bsy u_x,\nu),\quad
\bsy s_x=\Phi(\bsy s,\bsy w,\bsy w_x,\mu),\\
\end{aligned}
\end{equation}
where $\Phi(\bsy u,\bsy v,\bsy v_x,\mu)$ denotes the right hand
side of the B\"acklund transformation (\ref{vecSKdVBT}).

By the vector formula of superposition we mean a relation
of the form
\begin{equation}\label{df}
\bsy s=k_1\bsy u+k_2\bsy v+k_3\bsy w,
\end{equation}
where the coefficients $k_i$ depend on the scalar products
of the vectors $\bsy u,\bsy v$ and $\bsy w$.

Eliminating the vectors $\bsy u_x,\bsy w_x$ and $\bsy s_x$
from system (\ref{4bvec}), we obtain a relation of the form
\begin{equation}\label{FS-begin}
(\bsy u-\bsy s,\bsy v_x)\bsy A
+(\bsy w-\bsy s,\bsy v_x)\bsy B
+(\bsy v-\bsy s,\bsy v_x)\bsy C
+a\bsy v_x=0,
\end{equation}
where the vectors $\bsy A,\bsy B,\bsy C$ and the scalar $a$
do not depend on $\bsy v_x$.

It is natural to assume (this was suggested by V. Adler)
that this relation is an identity with respect to $\bsy v_x$.
This means that $a=0$ and that, replacing $\bsy v_x$ by
$\bsy u-\bsy s,\bsy w-\bsy s$, and $\bsy v-\bsy s,$
we obtain three vector relations, which do not depend on
the derivative $\bsy v_x$.
Their scalar products with the vector variables, together
with the relation $a=0,$ give us an algebraic system for
the scalar products
$(\bsy u,\bsy s),(\bsy v,\bsy s),(\bsy w,\bsy s),(\bsy s,\bsy s)$.

Non-trivial calculations show that this overdetermined system
has two solutions. One of them leads to the superposition
formula \cite{meshsok1}
\begin{equation}\label{sup}
\bsy s=\bsy v+(\mu-\nu)
\frac{
\nu(\bsy v-\bsy w)^2(\bsy v-\bsy u)
-\mu(\bsy v-\bsy u)^2(\bsy v-\bsy w)
}{
\big(\mu(\bsy v-\bsy u)-\nu(\bsy v-\bsy w)\big)^2
}.
\end{equation}
The second solution corresponds to the replacement
$\mu\mapsto-\mu.$

The corresponding Bianchi diagram
\begin{equation*}
\begin{CD}
{\bsy u} @>\nu>> {\bsy s}\\
@A\mu AA @AA\mu A\\
{\bsy v} @>>\nu> {\bsy w}
\end{CD}
\end{equation*}
admits the symmetry group generated by the reflection
$\bsy u\leftrightarrow\bsy w$ and $\pi/2$ rotation.
The symmetry of formula \eqref{sup} with respect to the
reflection (with $\mu\leftrightarrow\nu$) is obvious.
Calculations similar to those described at the end of
Section 3.1 show that the formula \eqref{sup} is symmetric
with respect to the rotation.

Note that \eqref{sup} can be rewritten as
$$
\mu\frac{\bsy w-\bsy v}{(\bsy w-\bsy v)^2}
-\nu\frac{\bsy u-\bsy v}{(\bsy u-\bsy v)^2}
=
(\mu-\nu)\frac{\bsy s-\bsy v}{(\bsy s-\bsy v)^2}.
$$
In this form it is a vector analog of the ``cross-ratio''
model \cite{ABobSur}. In the scalar limit the latter formula
reduces to (\ref{supfor}).

\section{Acknowledgements}

The authors thank V.E. Adler for insightful comments and
constant attention to the research and A.G. Meshkov for allowing
the authors to use the Maple package for operating with vector
differential equations, written by him.
The research of VVS is an output of a research project
implemented as part of the Basic Research Program at the
National Research University Higher School of Economics
(HSE University).

\end{document}